\documentclass[11 pt,a4paper]{article}
\usepackage{amsmath}
\usepackage[pdftex]{graphicx}
\usepackage{epsfig}
\usepackage{epstopdf}
\graphicspath{{pictures/}}
\usepackage{geometry}
\usepackage{cite}
\usepackage{amssymb}
\usepackage{amsfonts}
\usepackage{amsthm}
\DeclareUnicodeCharacter{200F}{}
\usepackage{hyperref}
\date{}

\usepackage{mathtools}
\usepackage{mathrsfs}
\usepackage{amsmath}

\newcommand{\bra}[1]{\bigl\langle #1 \bigr|}
\newcommand{\ket}[1]{\bigl| #1 \bigr\rangle}
\begin{document}
	\begin{center}
\textbf{\Large Exchanging  quantum correlations and non-local  information between three qubit-syatem\\}
	\bigskip	
F. Ebrahiam$^{a}$\textit{\footnote{fawzeyaebrahimali@gmail.com }}, N. Metwally$^{a,b}$\textit{\footnote{e-mail:nmetwally@gmail.com}},\\
$^{a}${\footnotesize Math. Dept., College of Science, University of Bahrain, Bahrain.}\\
$^{b}${\footnotesize Department of Mathematics, Aswan University
	Aswan, Sahari 81528, Egypt.}\\
	\end{center}

\begin{abstract}

The possibility of  exchanging the quantum correlations and the non-local information between three qubits interact directly or indirectly via Dzyaloshinskii-Moriya (DM)is discussed.  The  initial state settings and the interaction strength represent  control parameters on the exchanging phenomena.  The  non-local information that  encoded on the different partitions doesn't exceed the initial one. It is shown that, the ability of DM interaction to generate entanglement is larger than that displayed for the dipole interaction. The possibility of maximizing the quantum correlations between the three qubits increases as one increase the strength of interaction and starting with large initial quantum correlations. The long-lived quantum correlations could be achieved by controlling the strength of the dipole interaction.
	
	\end{abstract}

\textbf{Keywords}:Quantum correlation, non-local information, Dzyaloshinski-Moriya interaction, Concurrence

\section{Introduction}
Entanglement is the core idea of  modern theoretical physics \cite{TEM}, where it  plays an  important role in different applications of quantum information processing (QIP,  like quantum cryptography\cite{Gisin}, quantum teleportation\cite{Bennt}, quantum dense coding\cite{Schumacher}, and etc. Therefore, there is a need to be  generated and quantified, as well as keep it survival for a long time. Indeed, it has been generated between different objects; atoms \cite{knight} charged qubits \cite{Nori},  quantum dots \cite{Edo},  etc.  Dzyaloshinskii-Moriya (DM), \cite{DM} represents one of the most common interaction that has used  widely  to generate quantum  correlation between different systems. For example, the possibility of generating a thermal entanglement between  two qubits  Heisenberg chain in  the presence of the Dzyaloshinski-Moriya is discussed in \cite{Guo}.  The effect of DM interaction on the dynamics of a two-qutrit system is investigated by Jafarpour and Ashrafpour\cite{Ashrafpour}. Sharma and  Pandey\cite{Sharma} studied the entanglement dynamics between qubit -qutrit systems via DM interaction. The  influence of the anisotropic antisymmetric exchange interaction,
 on the entanglement of two qubits is examined in\cite{ZEYNEP}. The DM interaction is used to generated  entangled quantum network\cite{Metwally}.   The orthogonality speed of two-qubit state interacts locally with spin chain in the presence of Dzyaloshinsky–Moriya interaction is investigated by Kahla, et. al \cite{Kahla}. Therefore, we are  motivated to discuses  the phenomena of exchanging  the quantum correlations and  the non-local information between three qubits, where two of them are either classically or non-classically correlated. It is assumed that, by using DM interaction one of the correlated qubits interacts locally with a control qubit that is  initially prepared in a vacuum state. The effect of the initial state settings of the correlated qubits and the strength of DM interaction on the exchange process is investigated.

The outline of the paper is arranged as: In Sec.(2), we review the most  common quantifiers of quantum correlation; concurrence, entanglement  of formation and negativity. The qubits system and its evolution is described in Sec.(3). The exchange of quantum correlations between the marginal states of the three qubits is discussed in Sec.(4). In  Sec.(5), the behavior of the exchanged non-local information is  investigated. Finally, we summarized our results in  Sec.(6).

\section{Measure of Entanglement}
In this section, we review some important quantifiers of quantum correlations (entanglement). Among of these quantifiers are; concurrence \cite{Wotter}, entanglement of formation, and the negativity. It is well known that,    a quantum system that consist of two qubits  ${\ket{\psi_{12}}}$,  is called a separable or classically correlated if ${\ket{\psi_{12}}}={\ket{\psi_{1}}} \otimes {\ket{\psi_{2}}}$, otherwise the system is entangled (quantum correlated). In the following subsections, we review the three measures in details.
\subsection{Concurrence}
This measure represents one of the most simplest quantifiers of entanglement, it is called Wootteres Concurrence \cite{Wotter}, where it  satisfies all the criteria of the good  measures of entanglement.
Mathematically it is defined as,
\begin{equation}
\mathcal{C}=max\{0,\sqrt{\lambda_1}- \sqrt{\lambda_2}-\sqrt{\lambda_3}-\sqrt{\lambda_4}\},
\end{equation}
where $\lambda_1\geq\lambda_2\geq\lambda_3\geq\lambda_4$ ,
 ${\lambda_i}$ (i=1,2,3,4)  are  the eigenvalues of  the density operator $\Tilde{{\rho_{12}}}$, such that
   \begin{equation}
       \Tilde{{\rho_{12}}} = {\rho_{12}} (\sigma_y^1 \otimes \sigma_y^2) \rho_{12}^\ast ( \sigma_y^1 \otimes \sigma_y^2).
       \label{Conc}
   \end{equation}
The operator ${\rho_{12}^\ast}$ is the complex conjugate of the density operator  of the system ${\rho_{12}}$, and $\sigma_y=i(\ket{0}\bra{1}-\ket{1}\bra{0})$ is Pauli operator in $y$-direction.  For separable states the concurrence  $\mathcal{C}=0$ and for the maximum entangled state $\mathcal{C}=1$.

\subsection{Entanglement of formation}
One of the meaningful measurement of  entanglement is the entanglement of formation $(\mathcal{E}_F)$ \cite{Wotters}.
 For a  two qubits  system the $\mathcal{E}_F$ is defined as a function of the concurrence \cite{Wotter} as,
   \begin{equation}
       \mathcal{E}_F({\rho})
       = H\Bigl(\frac{1+{\sqrt{1-{{\mathcal{C}({\rho})}^2}}}}{2}\Bigr),
   \end{equation}
 where $H(x)=-x\;log_2x-\;(1-x)\;log_2(1-x)$ represents  the Shannon Entropy function and $0\leq \mathcal{E}_F \leq 1$.

\subsection{Negativity}
The third measure of entanglement is the negativity, which is based on the eigenvalues of the partial transpose quantum system \cite{Peres,Horodecki}. However  for two qubit  system $\rho_{12}$, the negativity  $\mathcal{N}$ is defined as,
\begin{equation}
   \mathcal{ N}=2\smashoperator{\sum_{i=1}^{4}} max(0,-{\mu_i}),
\end{equation}
where ${\mu_i}$ are the negative eigenvalues of the partial transpose of the given quantum system, namely $\rho_{ab}^{T_{1(2)}}$.  Similarly the negativity $\mathcal{N}({\rho_{12}})$ is  ranged between $0$ and $1$, where  $\mathcal{N}({\rho})$=0 if the quantum system is separable and $\mathcal{N}({\rho})$=1 if the given quantum system is maximally entangled.
\section{The suggested model}
The system composed of two qubits $A$, and $B$, are prepared in a partial entangled state interacts by using Heisenberg XX Spin model \cite{BR}. It is assumed that, one of the  subsystems, say $A$, interacts with   third qubit $C$, as a  controller via  Dzyaloshinskii-Moriya $(DM)$. The total  Hamiltonian which describes this system is given by,
\begin{equation}
   \mathcal{H}_{sys}=\frac{1}{2}\omega\bigl(\sigma_A^x \sigma_B^x +  \sigma_A^y \sigma_B^y\Bigr)+\boldsymbol{D}\cdot(\boldsymbol{\sigma_{A}\times\sigma_C)},
\end{equation}
where $\boldsymbol{D}=(D_x,D_y, D_z)$ is  the DM interaction, and $D_i, i=x,y,z$ are the interaction strength on the three directions.
$\omega$ is the coupling parameter between qubits $A$, $B$, and
$ \sigma_i^x=\begin{pmatrix*}[r] 0 & 1 \\ 1 & 0 \end{pmatrix*}$, $ \sigma_i^y=\begin{pmatrix*}[r] 0 & \imath \\ -\imath & 0 \end{pmatrix*}, i=A, B$ are the Pauli operators for both qubits.
In this contribution, we  assume that  $\boldsymbol{D}$ to be  polarized on the $z$ direction, namely, $\boldsymbol{D} = ( 0 , 0 , D_z )$.
Let us assume that the  initial state of the whole system at time t=0 is given by,
\begin{equation}
   \rho_{s}(0)=\rho^{AB}(0)\otimes\rho^C(0),
   \end{equation}
where
\begin{equation}
\rho^{AB}(0)=\kappa\ket{\varphi^{ab}}\bra{\varphi^{ab}}+\frac{1}{4}(1-\kappa)I_{4\times 4}, ~\rho^c(0)=\ket{\varphi^c}\bra{\varphi^c},
\end{equation}
 with $\ket{\varphi^{AB}}=\cos(\alpha)\ket{eg}+\sin(\alpha)\ket{ge}$ and  $\ket{\varphi^{c}}=\cos(\gamma)\ket{e}+\sin(\gamma)\ket{g}$,
 $I_{4\times 4}$ is the identity operator. It is clear that, at $\alpha=\frac{\pi}{4}$, the state $\ket{\varphi^{AB}}$  reduces to Bell state  $\ket{\psi^{+}}$ .
At $t>0$, the  total density operator  of the whole system is given by,
\begin{equation}
    \rho_s(t)\;=\mathcal{U}(t)\rho_s(0)\mathcal{U}(t)^\dagger,~ \mathcal{U}(t)=\exp[-i\mathcal{H}_{sys} t],
\end{equation}
where the unitary operator $\mathcal{U}(t)$ takes the form,
\begin{equation}
\mathcal{U}(t)=\sum_{i=1}^2{\alpha_\cdot\beta_i},
\label{unitary}
\end{equation}
where,
\begin{eqnarray}
    \alpha_1&=&\cos\bigl(\;\frac{w}{2}\;t\;\bigr)\;.\;I_{8\times 8} -\;\imath\sin(\;\frac{w}{2}\;t\;)\;\sigma_A^x\;\otimes\; \sigma_B^x\;\otimes\;I_{2\times 2}
    \nonumber\\
    \alpha_2&=&\cos(D_z t\;)I_{8\times 8} -\;\imath\sin(D_Z\;t\;)\;\sigma_A^x\;\otimes\;I_{2\times2}\; \otimes\;\sigma_C^y,
    \nonumber\\
    \beta_1&=&\cos(\;\frac{w}{2}\;t\;)\;.\;I_{8\times 8} -\;\imath\;sin(\;\frac{w}{2}\;t\;)\;\sigma_A^y\;\otimes\; \sigma_B^y\;\otimes\;I_{2\times 2},
    \nonumber\\
    \beta_2&=&\;cos(\;D\;t\;)\;.\;I_{8\times 8} +\;\imath\;\sin(D_z t\;)\;\sigma_A^y\;\otimes\;I_{2\times 2}\;\otimes\; \sigma_C^x.
\end{eqnarray}
 Then by using the unitary operator (\ref{unitary}) and the initial state $\rho_s(0)$, one gets the final state $\rho_s(t)$ at any $t>0$. Due to this interaction, there well be new entangled states are generated; $\rho_{AC}$ is generated between the qubits $A$, $C$, and  $\rho_{BC}$ between the qubits $B$ and $C$. Moreover, the amount of entanglement of  the initial state $\rho_{AB}$ will be affected by this interaction.
  In this context, it is important to quantify the amount of entanglement  between the three partitions $AB, AC,$ and $BC$, which are defined by the final states $\rho_{AB}(t), \rho_{AC}(t)$ and $\rho_{Bc}(t)$, respectively.

In the computational basis the three final density operators are given by a matrix of  size $4\times 4$. The non-zero elements  of the states $\rho_{k\ell}^{ij}$, $ij=ab,ac, bc$, respectively, and $k\ell=AB, Ac, BC$ are given by,

\begin{equation}
    \rho_{ij}(t)=\begin{pmatrix*}[r]
    \rho_{00}^{ij} & 0 & 0 & 0 \\
    0 & \rho_{11}^{ij} & \rho_{12}^{ij} & 0 \\
    0 & \rho_{21}^{ij} & \rho_{22}^{ij} & 0 \\
    0 & 0 & 0 & \rho_{33}^{ij} \\
    \end{pmatrix*}.
\end{equation}
The density operator of the state $\rho_{ab}(t)$ is defined by the following elements:
\begin{eqnarray}
\rho_{00}^{ab}&=&\frac{1}{4}(\kappa)\cos^2(2 D_z t),
\nonumber\\
\rho_{11}^{ab}&=&\frac{1}{8}(2\cos^2(2t)\cos^2(2 D_z\;t)+(3+3\kappa-(1-\kappa)\cos(4 D_z t))\sin^2(2t)),
\nonumber\\
\rho_{22}^{ab}&=&\frac{1}{8}(2\sin^2(2t)\cos^2(2 D_z t)+(3+3\kappa-(1-\kappa)\cos(4D\;t))\cos^2(2t)),
\nonumber\\
\rho_{21}^{ab}&=&\frac{1}{8}(2\sqrt{3}\kappa\cos(2 D_z t))-\imath\sin(4t)\;(-\frac{1}{2}\kappa(3+cos(4 D_z t)-2sin^2(2 D_z\;t))),
\nonumber\\
\rho_{33}^{ab}&=&\frac{1}{4}(1-\kappa+\sin^2(2 D_z t)), ~~\rho_{12}^{ab}=\rho_{21}^{ab*}.
\end{eqnarray}
Similarly the state $\rho_{ac}$ is described  by the following non-zero elements,
\begin{eqnarray}
\rho_{00}^{ac}&=&\frac{1}{4}(1-\kappa)\sin^2(2t)\sin^2(2 D_z\;t),
\nonumber\\
\rho_{11}^{ac}&=&\frac{1}{4}((1-\kappa+\cos^2(2t))\;cos^2(2\;D_z\;t)+(1+2\kappa)\sin^2(2t)),
\nonumber\\
\rho_{22}^{ac}&=&\frac{1}{8}(2\;\imath\;\sqrt{3}\kappa\sin(2t)\sin(2\;D_z t)+(\kappa-2)\cos(2t)\sin(4 D_z\;t)),
\nonumber\\
\rho_{21}^{ac}&=&-\frac{1}{8}(\kappa-3-(1-\kappa)\cos(4\;t))\sin^2(2\;D_z t),
\nonumber\\
\rho_{33}^{ac}&=&\frac{1}{16}(7+(1+4\kappa)\cos(4t)+8\cos^2(t)\cos(4 D_z t)\sin^2(t)), ~\rho_{12}^{ac}=\rho_{21}^{ac*}.
\end{eqnarray}
Finally the non zero elements of  third partition $\rho_{bc}$ are given by,
\begin{eqnarray}
 \rho_{00}^{bc}&=&\frac{1}{4}(1-\kappa)\cos^2(2t)\;sin^2(2\;D\;t),
 \nonumber\\
 \rho_{11}^{bc}&=&\frac{1}{4}((1-\kappa+\sin^2(2t))\;cos^2(2\;D\;t)+(1+2\kappa)\cos^2(2t)),
 \nonumber\\
\rho_{22}^{bc}&=&\frac{1}{8}\;\kappa\sin(2 D_z\;t)(-2\;\sqrt{3}\cos(2t)+\;\imath\;(\sin(2(1+D_z)t)+sin(2(1-D_z)t)\;),
\nonumber\\
\rho_{21}^{bc}&=&\frac{1}{4}(1+(1-\kappa)\;sin^2(2\;t))\;sin^2(2\;D_z\;t),
\nonumber\\
\rho_{33}^{ac}&=&\frac{1}{8}(3-(1+2\kappa)\;cos(4t)+2\;cos^2(2t)\;cos(2\;D_z\;t), ~\rho_{12}^{bc}=\rho_{21}^{bc*}.
\end{eqnarray}

\section{Exchanging  quantum correlations }
\begin{figure}[h!]
	\centering
	\includegraphics[width=0.4\linewidth, height=5cm]{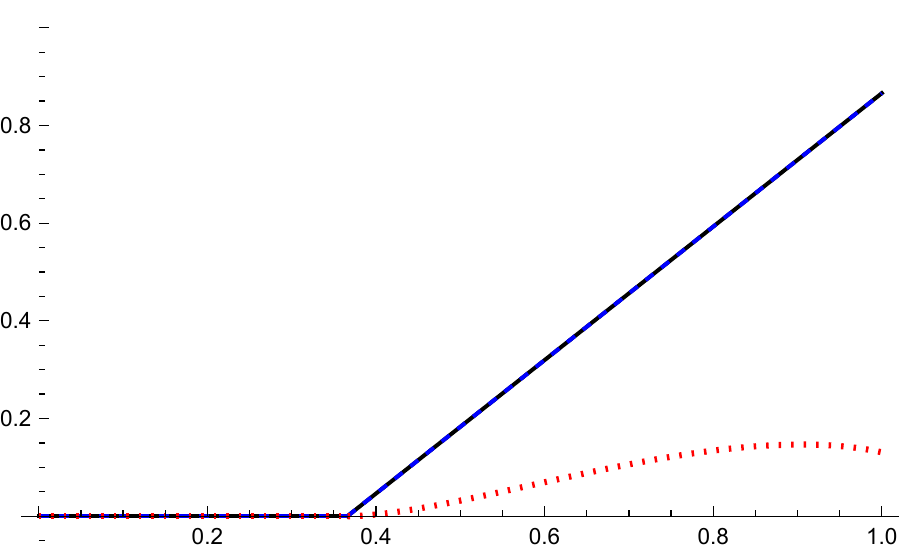}~~\quad
\put(-190,50){\rotatebox{90}{\small{$Q_c$-quantifiers}}}
 \put(-70,-10){$\kappa$}
		\caption{The amount of quantum correlation of the initial state $\rho_{ab}(0)$ that predicted via concurrence, entanglement of formation and negativity.}
	\label{Fig1}
\end{figure}

In this section we investigate the amount of entanglement that loses from the initial state $\rho_{ab}(0)$ and that exchanged  between the terminals of the initial state and the control  qubit $C$.
 In Fig.(\ref{Fig1}), we quantify the amount of entanglement on the initial state via the concurrence, $\mathcal{C}$, negativity, $\mathcal{N}$ and the entangkement of formation $\mathcal{E}_F$. It it is assumed that, the initial state of the system is prepared by setting $\alpha=\pi/3$, namely $\rho_{ab}$   is a partial entangled state.  It is clear that, the quantum correlation between the subsystems $"a"$ and $"b"$ appears  at larger values of the weight parameter $\kappa$, namely $\kappa>0.37$. From this figure, it is clear that, the predicted quantum correlation depends on the used  quantifier. However, the concurrence and the negativity  predict the entanglement at smaller values of the weight $\kappa$. Moreover, the smallest upper bounds of the quantum correlations are predicted by using the entanglement of formation, $\mathcal{E}_F$.

\subsection{Lose and gain quantum  correlations of  $\rho_{AB}$}
Due to the interaction with the environment which is described by the control qubit $C$, the initial state $\rho_{ab}$ loses some of its quantum correlation. Therefore, it is important to investigate the effect of interaction's parameters on the  quantum correlation of the initial quantum system.

\begin{figure}[h!]
	\centering
    \includegraphics[width=0.32\linewidth, height=4cm]{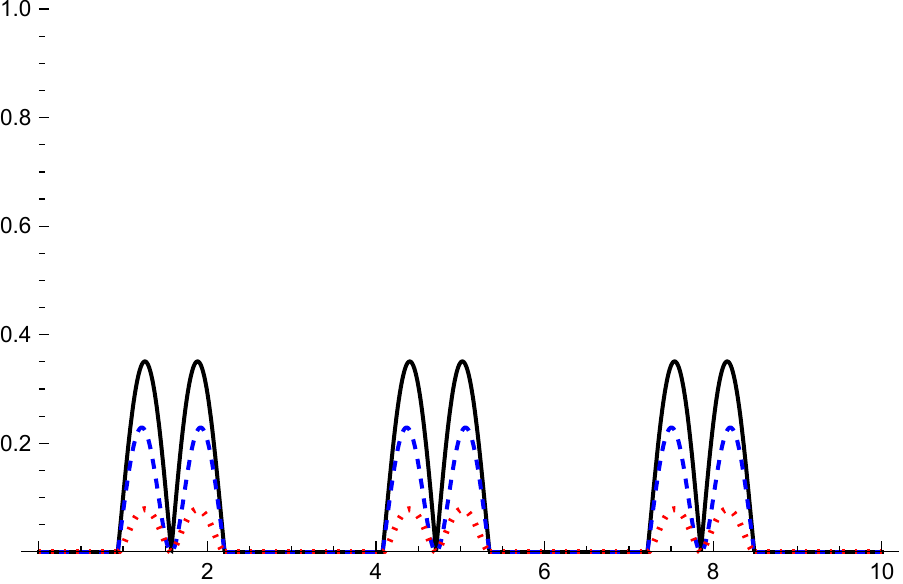}~~\quad\quad
    \put(-70,-10){$t$}
\put(-160,50){\rotatebox{90}{\small{$Q_c$-quantifiers}}}
    \put(-50, 100){$(a)$}
    \hspace{1.5cm}
	\includegraphics[width=0.32\linewidth, height=4cm]{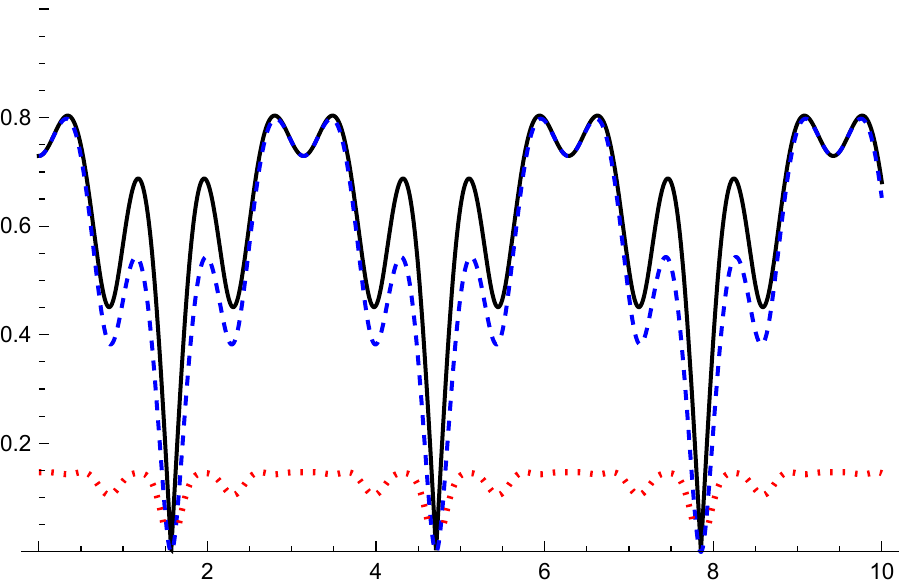}
 \put(-70,-10){$t$}
    \put(-160,50){\rotatebox{90}{\small{$Q_c$-quantifiers}}}
    \put(-60, 100){$(b)$}
    	\caption{The dynamics of the  quantum correlation loses of the  state $\rho_{ab}(t)$ that predicted by using the concurrence (solid), the negativity (dash), and the entanglement of  formation (dot-lines). It is assumed that, $\rho_{ab}(0)$ is prepared such that $\alpha=\pi/3$ and control qubit is prepared by setting $\gamma=\pi/2$, while $D_z=0.5$ (a) $\kappa=0.3$ and (b)$\kappa=0.9$.}
	\label{Fig2}
\end{figure}
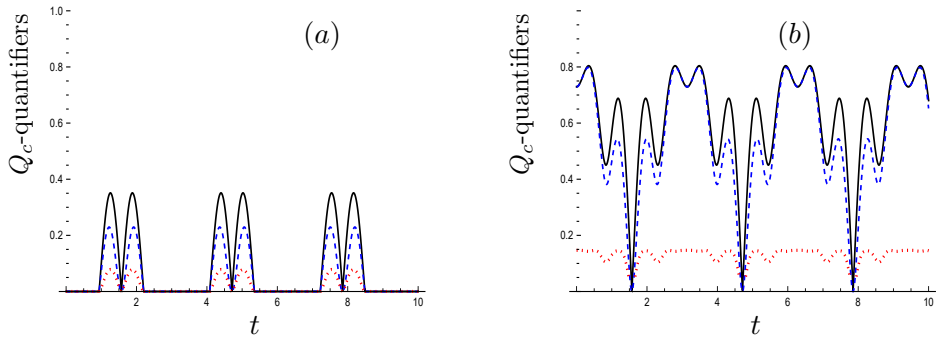
Fig.(\ref{Fig2}) shows the behavior of the three quantifiers for a system is initially prepared   in a  partial entangled state, where $\alpha=\pi/3$ and different values of the weight parameter $\kappa$. Moreover, the control qubit is prepared in the state $\rho_c=\ket{0}\bra{0}$, namely we set $\gamma=\pi/2$.  From Fig.(\ref{Fig2}a), it is clear that, at small value of the weight parameter, $\kappa=0.3$, the quantum correlation is generated at $t\simeq 0.9$. However, the quantifiers increase gradually to reach their maximum bounds for the first time at $t=1.2$ and decrease to vanish completely at $t\simeq 1.6$. The largest bounds are predicted for the concurrence $\mathcal{C}$, while the smallest bounds of quantum correlations are predicted by the entanglement of formation $\mathcal{E}_F$.

In Fig.(\ref{Fig2}b), we investigate the behavior of the quantum correlation  at large value  the weight parameter $\kappa$. It is clear that, at  $t=0$, all the quantifiers predicate the presence of the quantum correlation. However, as the interaction time increases, the predicted  quantum correlations flocculated between their lower and upper bounds. Moreover, the lower bounds that displayed by the concurrence $\mathcal{C}$ are the largest one, but the maximum values of $\mathcal{C}$ and the negativity $\mathcal{N}$ are coincide.  As it is shown from Fig.(\ref{Fig2}b), the smallest amount of quantum correlation are
 quantified by using the $\mathcal{E}_F$.

 From Fig.(\ref{Fig2}), one may conclude that, the weight of the initial state plays a central role on keeping the  existence of the quantum correlations. If the weight parameter is small enough such that the initial state behaves as a product sate, the  DM interaction has the strength to generate quantum correlation between the subsystems of the initial state.
\begin{figure}[h!]
	\centering
    \includegraphics[width=0.4\linewidth, height=5cm]{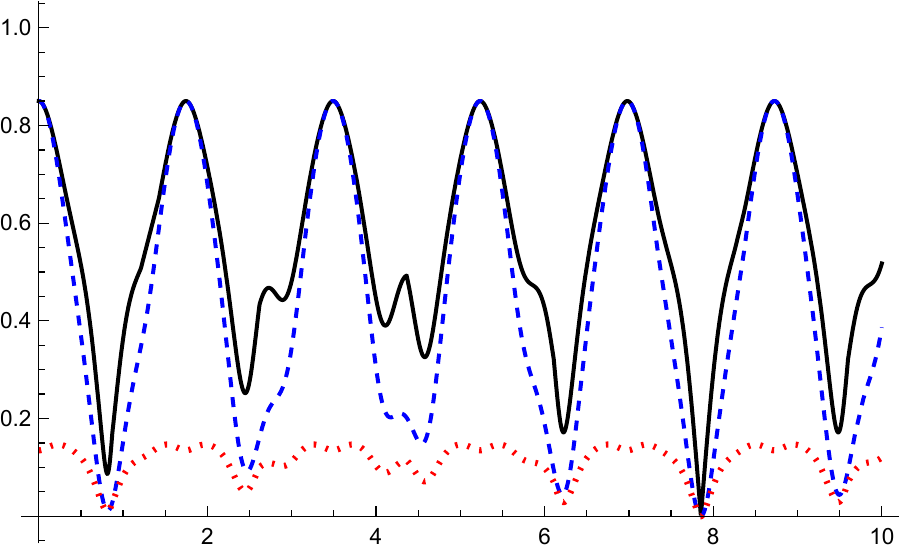}~~\quad\quad
    \put(-80,-10){$t$}
\put(-200,50){\rotatebox{90}{\small{$Q_c$-quantifiers}}}
	\caption{The same as Fig(\ref{Fig2}b), but $D_z=0.9$.}
	\label{Fig3}
\end{figure}

To clarify the role that played by the strength of DM interaction, we investigate in Fig.(\ref{Fig3})  the behavior of the three quantifiers at larger  values of  DM interaction, where we set $D_z=0.9$.  The predicted behavior of the three quantifiers  shows that, as soon as the interaction is switched on, the quantum correlations that predicted  decrease suddenly to reach their  minimum values. At further interaction time, the correlation rebirths  again to reach their maximum values.  The maximum values of the concurrence and the negativity are coincide, while the lower bounds of $\mathcal{C}$ are much better than that displayed by  the negativity $\mathcal{N}$. The minimum and maximum values of the quantum correlations that predicted via the three quantifiers are displayed at the same interaction time.

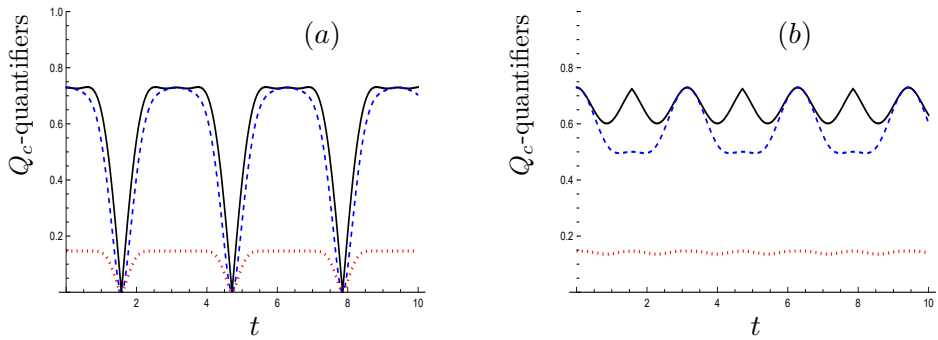
\begin{figure}[h!]
	\centering
    \includegraphics[width=0.32\linewidth, height=4cm]{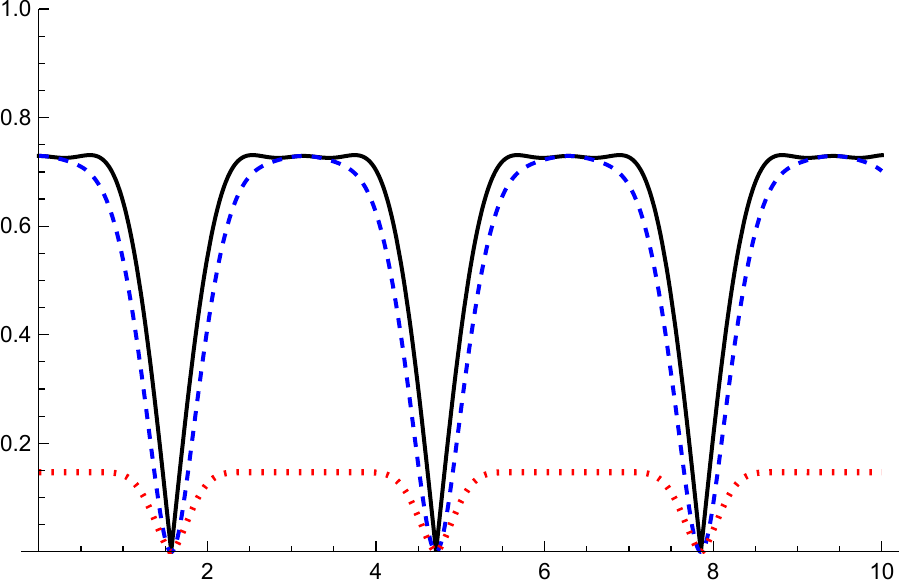}~~\quad\quad
    \put(-70,-10){$t$}
\put(-160,50){\rotatebox{90}{\small{$Q_c$-quantifiers}}}
    \put(-50, 100){$(a)$}
    \hspace{1.5cm}
	\includegraphics[width=0.32\linewidth, height=4cm]{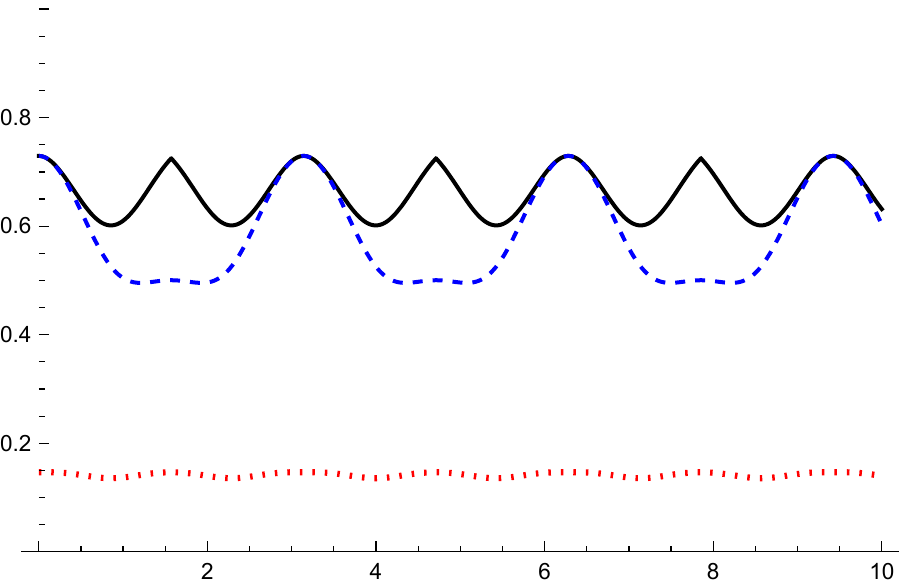}
 \put(-70,-10){$t$}
    \put(-160,50){\rotatebox{90}{\small{$Q_c$-quantifiers}}}
    \put(-60, 100){$(b)$}
	\caption{The same as Fig.(\ref{Fig2}) but (a) $\omega=1$, and (b) $\omega=0.5$. The initial and the control states are prepared such that $\alpha=\pi/3, \kappa=0.9$ and $\gamma=\pi/2$.}
	\label{Fig4}
\end{figure}
Fig.(\ref{Fig4}), exhibits the effect of the dipole's strength interaction  where we  set $\omega=0.5,1$, while all the parameters values are the same as Fig.(\ref{Fig2}b). The predicted behavior is similar to that displayed in Fig.(\ref{Fig3}). As it is shown  from Fig.(\ref{Fig4}a), where $\omega=1$, all the quantifiers predict a long-lived quantum correlations. The phenomena of the sudden death/birth are predicted at the same  interaction time for all the quantifiers.  Moreover, the quantum correlations vanish at larger interaction time compared with that displayed in Fig.(\ref{Fig3}).  As one increases the strength of the dipole interaction,($\omega=0.5$) in Fig.(\ref{Fig4}b), the long-lived behavior of quantum correlation is lost and the oscillation behavior is  shown. Moreover, the minimum values of these oscillations are larger than those displayed in Fig.(\ref{Fig3}).

From the results that  displayed in Fig.(\ref{Fig4}), one can  consider the  strength of the dipole interaction as a control parameter to generated a long-lived quantum correlations between the terminals of the initial quantum system. Moreover,  it could be used to improve the lower bounds of these correlations.

\subsection{Lose and gain quantum  correlations of $\rho_{AC}$}

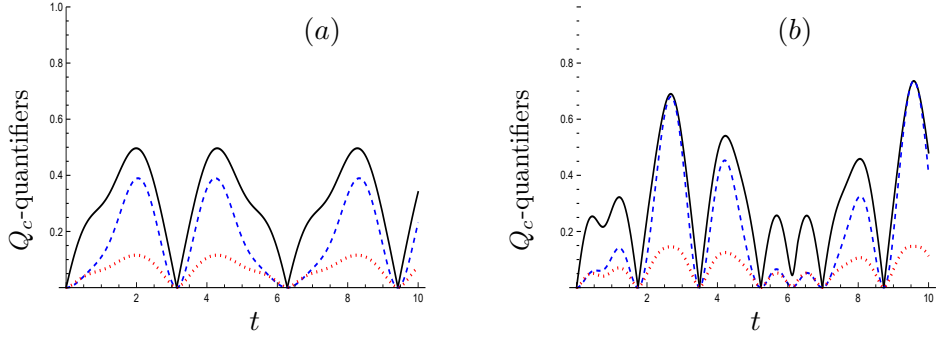
\begin{figure}[h!]
	\centering
    \includegraphics[width=0.32\linewidth, height=4cm]{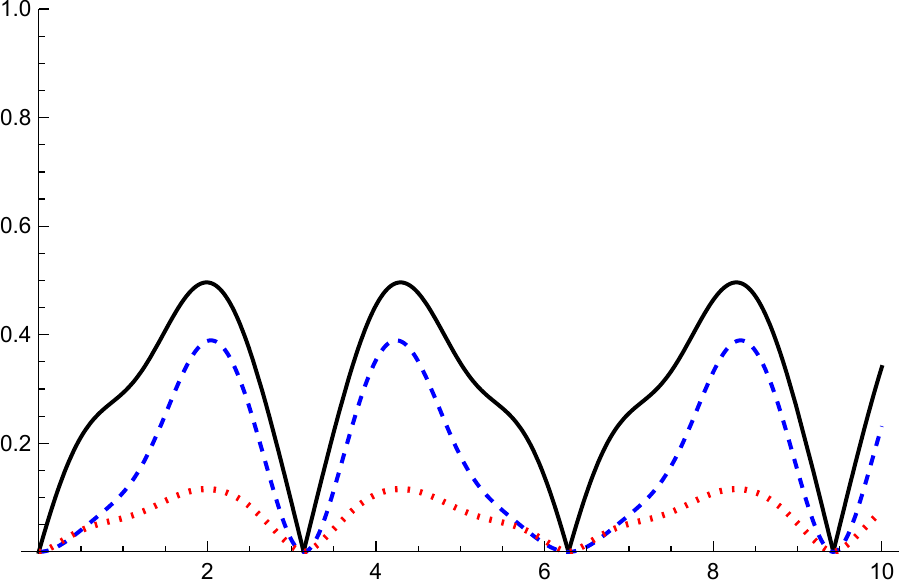}~~\quad\quad
    \put(-70,-10){$t$}
\put(-160,20){\rotatebox{90}{\small{$Q_c$-quantifiers}}}
    \put(-50, 100){$(a)$}
    \hspace{1.5cm}
     \includegraphics[width=0.32\linewidth, height=4cm]{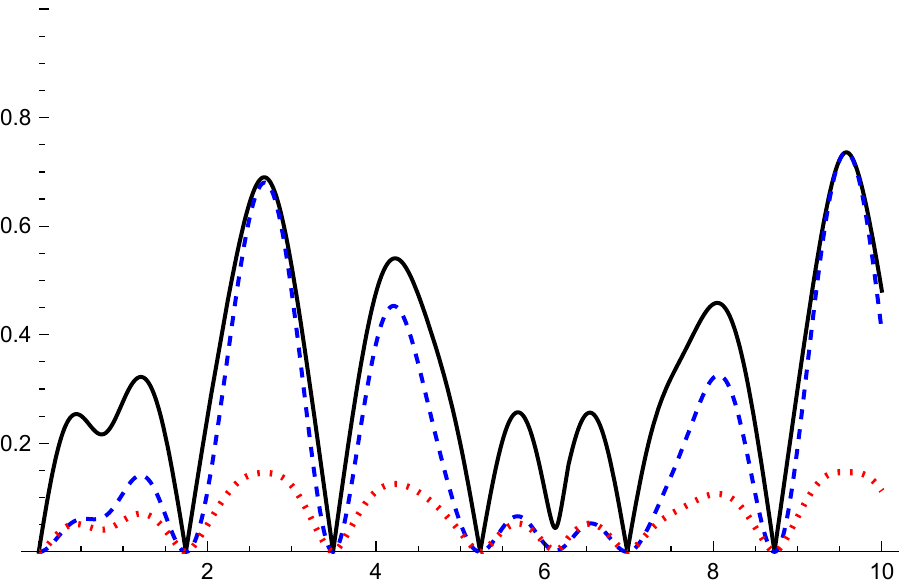}
 \put(-70,-10){$t$}
    \put(-160,20){\rotatebox{90}{\small{$Q_c$-quantifiers}}}
    \put(-60, 100){$(b)$}
    	\caption{The amount of  quantum correlation that is generated between the subsystems of the state $\rho_{AC}$, $\kappa=0.9, \omega=0.5, \gamma=\pi/2$, $\alpha=\pi/3$ and(a) $D_z=0.5$(b) $D_z=0.9$}
	\label{Fig5}
\end{figure}
The amount of quantum correlation that is generated between the subsystems $A$ and $C$ is described in Fig.(\ref{Fig5}), where the same initial state settings are considered. It is clear that, as soon as the interaction is switched on between the subsystem $A$ and the control parameter $C$, an entangled state is generated. All the quantifiers $\mathcal{C}, \mathcal{N}$ and $\mathcal{E}_{f}$ predict  that there is a quantum correlation is generated between the two qubits. From Figs.(\ref{Fig2}a) and (\ref{Fig5}a), it is clear DM interaction generates an entangled state between the subsystems $A$ and $C$, while the qubits $A$ and $B$ are disentangled at small values of $\kappa$. This means that, the ability of generating entanglement via DM is stronger than that may be generated by using dipole interaction. On the other hand, at large weight, namely $\kappa=0.9$, the quantum correlation is generated  between the qubits $A$ and $C$ on the expanse of the initial quantum correlation between $A$ and $B$. This phenomena is displayed in Fig.(\ref{Fig5}b), by investigating the behaviors of the three quantifiers, where as soon as the interaction is switched on, the quantum correlation of the state $\rho_{AB}$ decreases, while it increases between the qubits $A$ and $C$.

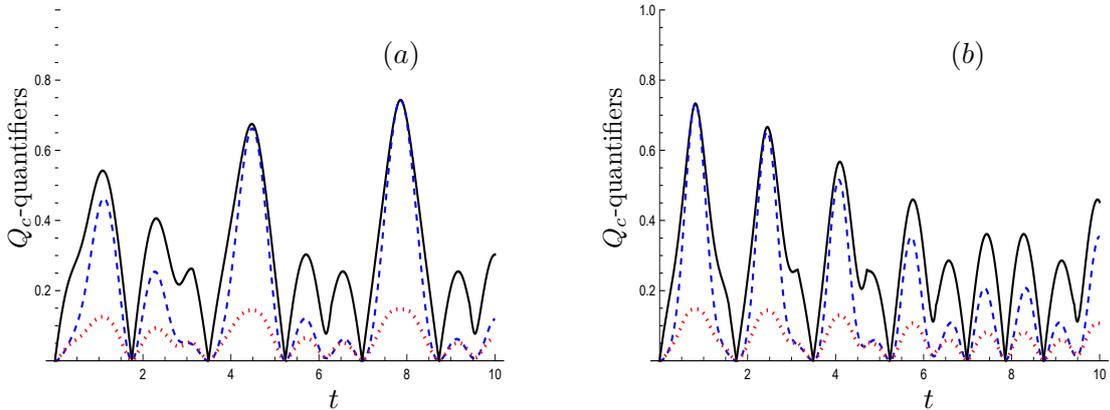
\begin{figure}[h!]
	\centering
    \includegraphics[width=0.4\linewidth, height=5cm]{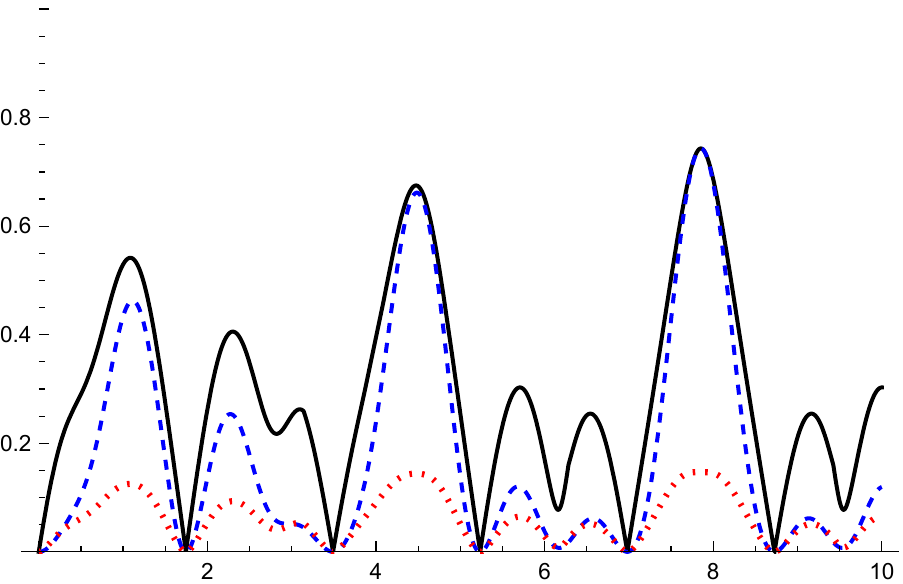}~~\quad\quad
    \put(-70,-10){$t$}
\put(-190,50){\rotatebox{90}{\small{$Q_c$-quantifiers}}}
    \put(-50, 120){$(a)$}
    \hspace{1.5cm}
	\includegraphics[width=0.4\linewidth, height=5cm]{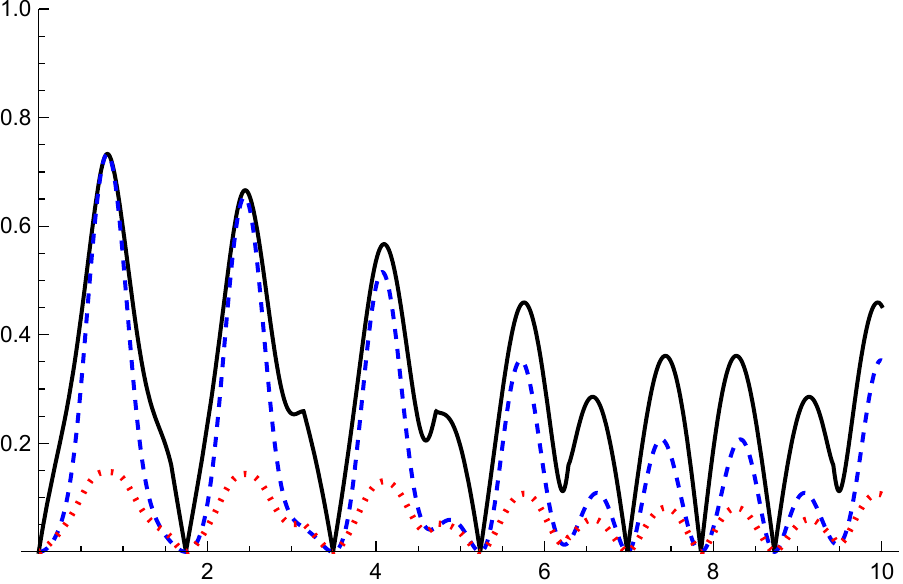}
 \put(-70,-10){$t$}
    \put(-190,50){\rotatebox{90}{\small{$Q_c$-quantifiers}}}
    \put(-60, 120){$(b)$}
	\caption{The same as Fig.(\ref{Fig4}) but  for the state $\rho_{AC}$(a) $\omega=1$, and (b) $\omega=2$. }
	\label{Fig6}
\end{figure}
In Fig.(\ref{Fig6}), we investigate the effect of the dipole interaction parameter on the  generated entangled state between the terminal $(A)$ and the control  qubit and $C$. To get a clear vision on this effect we consider different values of the dipole parameter, where we set $\omega=1,2$. It is clear from Fig.(\ref{Fig6}a), at  small values of the dipole interaction, means that less  quantum correlation between the initial qubits $A$ and $B$,  a large quantum correlation between the qubits $A$ and $C$  is predicted. Moreover,  small values of $\omega$ increases the  survival time of the quantum correlation of the state $\rho_{AC}$. Fig.(\ref{Fig6}b) displays the behavior of quantifiers at large values of the dipole interaction, where we set  $\omega=2$. The behavior is similar to that displayed in Fig.(\ref{Fig6}a), but the upper bounds of all the quantifiers are larger than those displayed at $\omega=1$.

\subsection{Lose and gain quantum  correlations of $\rho_{BC}$}
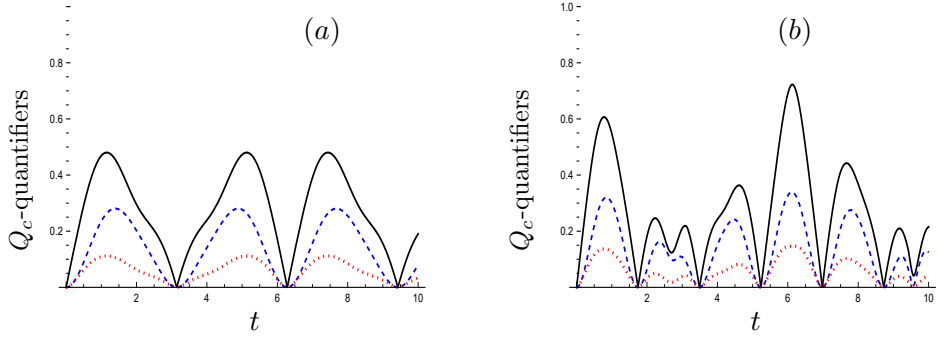
\begin{figure}[h!]
	\centering
    \includegraphics[width=0.32\linewidth, height=4cm]{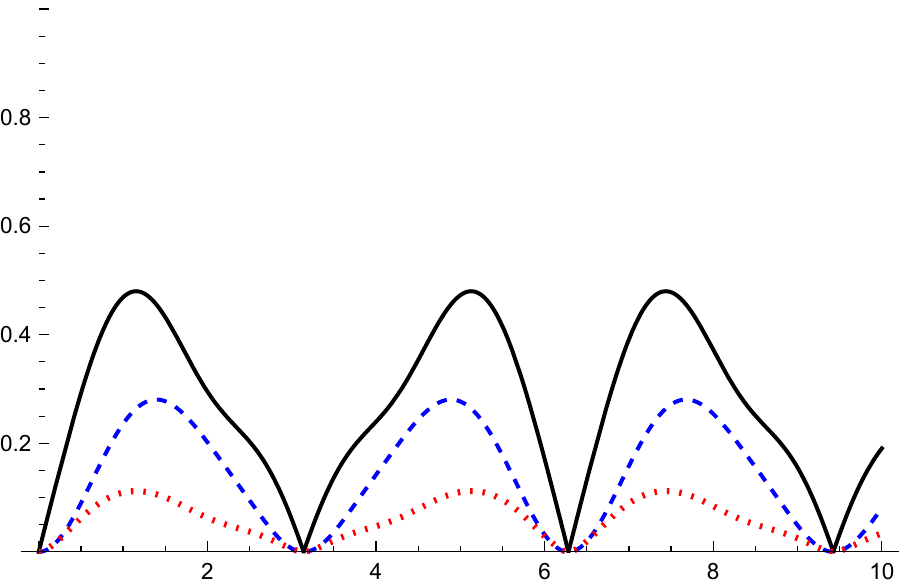}~~\quad\quad
    \put(-70,-10){$t$}
\put(-160,20){\rotatebox{90}{\small{$Q_c$-quantifiers}}}
    \put(-50, 100){$(a)$}
    \hspace{1.5cm}
	\includegraphics[width=0.32\linewidth, height=4cm]{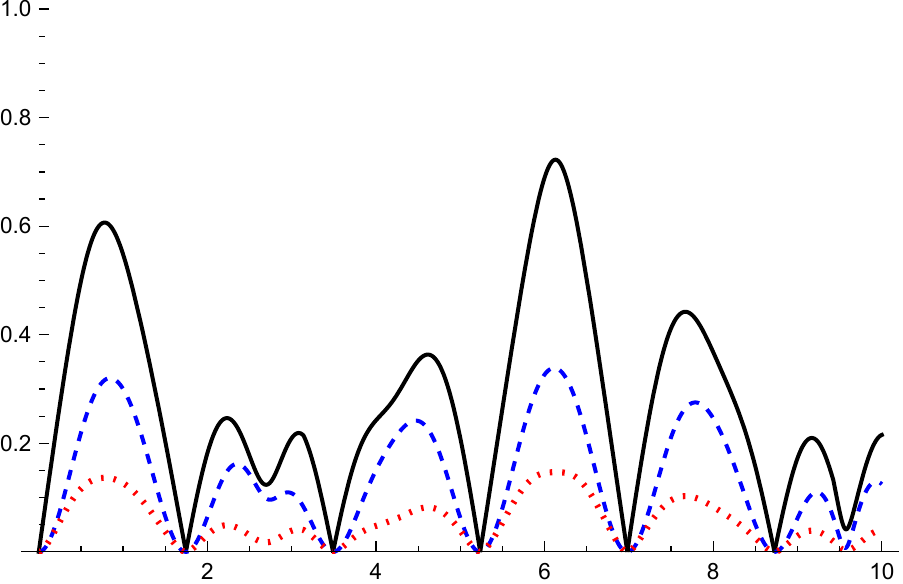}
 \put(-70,-10){$t$}
    \put(-160,20){\rotatebox{90}{\small{$Q_c$-quantifiers}}}
    \put(-60, 100){$(b)$}
    	\caption{The  same as Fig.(\ref{Fig5}) but for the state $\rho_{BC}$ .}
	\label{Fig7}
\end{figure}

As mentioned above, the DM interaction generates an entangled  between the qubits $B$ and $C$ represented by the density operator $\rho_{BC}$.  The behavior of the three quantifiers of the entanglement is predicted in Fig.(\ref{Fig7}) at different values of the interaction strength.
The behavior of the three quantifiers for the marginal state $\rho_{BC}$ at $D_z=0.5$ is displayed in Fig.(\ref{Fig7}a), where it displays a similar behavior  to that shown  in Fig.(\ref{Fig5}a) for the marginal state $\rho_{AC}$. However, the generated quantum correlation the  depicted by the three quantifiers increases suddenly as soon as the interaction is switched on, while the gradual behavior is displayed for the marginal state $\rho_{AC}$.  Moreover, the maximum values of the quantum correlations that shown by the three quantifiers are predicted at smaller interaction time compared with that displayed for the marginal state $\rho_{AC}$.  As  it is displayed from  Fig.(\ref{Fig7}b), the maximum bounds of quantum correlations are larger than those displayed in Fig.(\ref{Fig7}a), where we increase the interaction' strength.   For both marginal states, the sudden birth/death of the quantum correlation are depicted at the same interaction time, meanwhile the maximum quantum correlations are displayed for marginal state $\rho_{AB}$.

\begin{figure}[h!]
	\centering
    \includegraphics[width=0.4\linewidth, height=5cm]{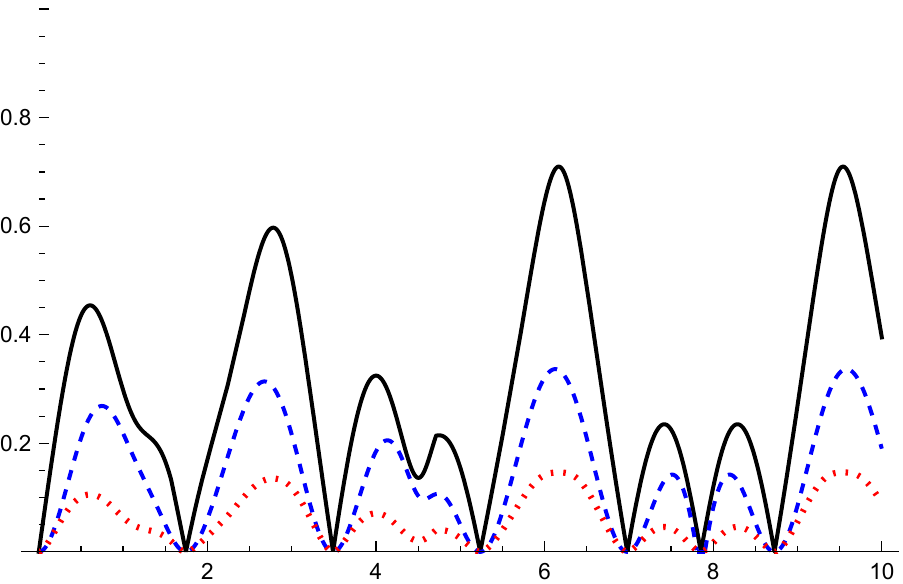}~~\quad\quad
    \put(-70,-10){$t$}
\put(-190,50){\rotatebox{90}{\small{$Q_c$-quantifiers}}}
    \put(-50, 100){$(a)$}
    \hspace{1.5cm}
	\includegraphics[width=0.4\linewidth, height=5cm]{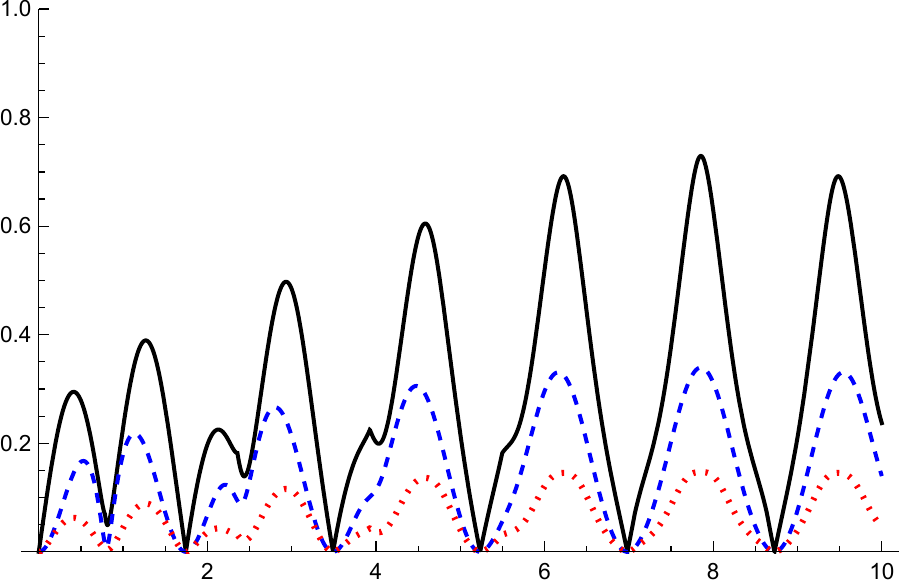}
 \put(-70,-10){$t$}
    \put(-190,50){\rotatebox{90}{\small{$Q_c$-quantifiers}}}
    \put(-60, 100){$(b)$}
	\caption{The same as Fig.(\ref{Fig4}) but  for the state $\rho_{BC}$(a) $\omega=1$, and (b) $\omega=2$. }
	\label{Fig8}
\end{figure}
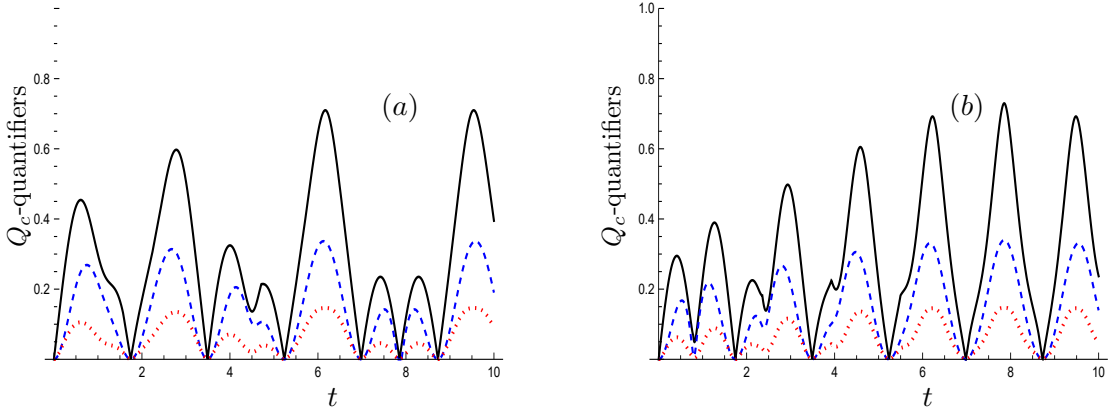
In Fig.(\ref{Fig8}), we examine the effect of the dipole interaction on the behavior of quantum correlation for the marginal state $\rho_{BC}$, where we set $\omega=1,2$ for Figs.(\ref{Fig8}a) and (\ref{Fig8}b), respectively.  It is clear that, the number oscillations of the three quantifiers are larger than those displayed at small value of $\omega$ as displayed in Fig.(\ref{Fig5}). Similarly, the phenomena of the sudden death/birth are displayed at smaller interaction time compared with those displayed in Fig.(\ref{Fig5}).  Moreover, the amount of the predicted quantum corrrelation for the marginal state $\rho_{AC}$, which is generated via direct interaction with DM, is larger than that displayed for the state $\rho_{BC}$, which is generated indirectly.

\section{Exchanging the Non-local information}
\begin{figure}[h!]
	\centering
	\includegraphics[width=0.4\linewidth, height=5cm]{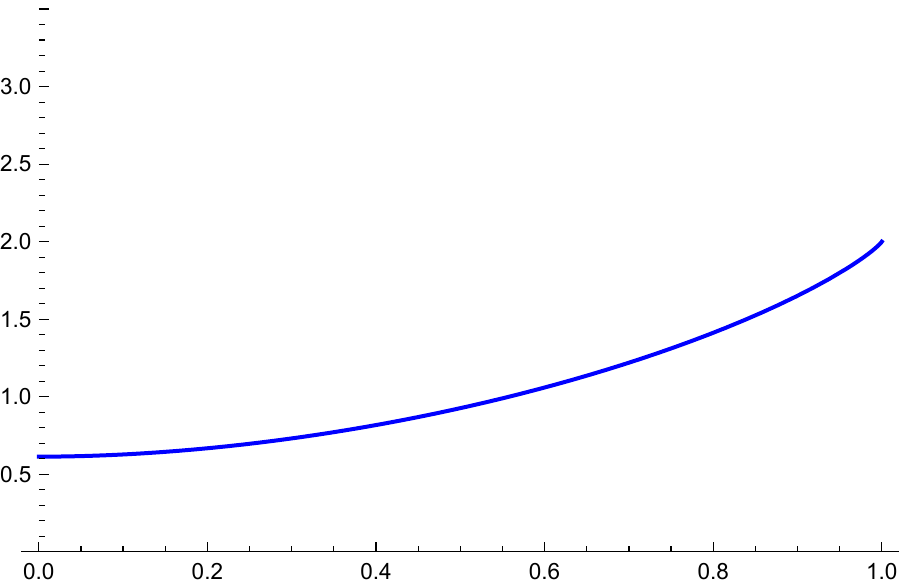}~~\quad
   \put(-70,-10){$\kappa$}
   \put(-50, 130){$(a)$}
\put(-200,50){\rotatebox{90}{\small{$I_{non}(\rho_{AB}(0))$}}}~~\quad
 \hspace{1.5cm}
\includegraphics[width=0.4\linewidth, height=5cm]{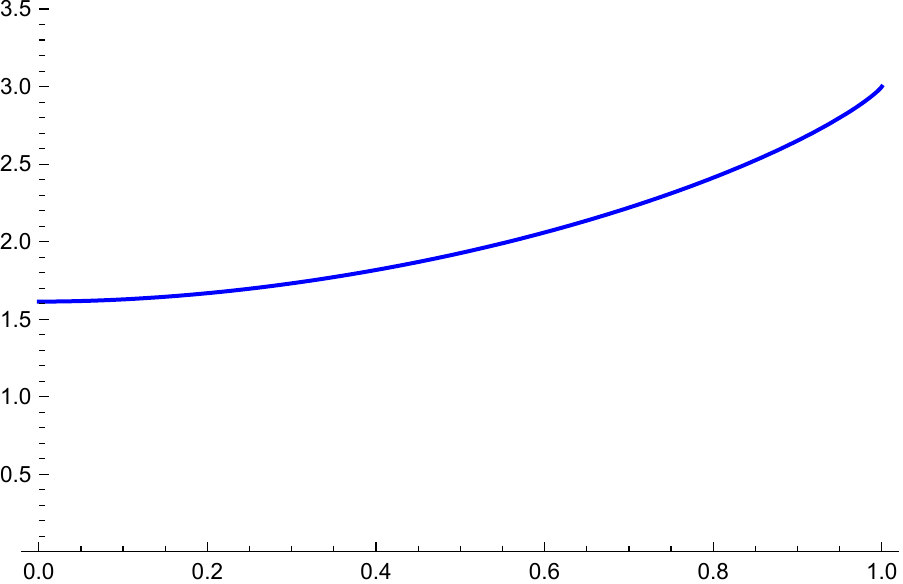}
 \put(-70,-10){$\kappa$}
    \put(-190,50){\rotatebox{90}{\small{$I_{non}(\rho_{ABC})$}}}
    \put(-60, 130){$(b)$}
		\caption{The non local information that coded on the states  (a)$\rho_{ab}(0)$and (b)  $\rho_{ABC}$, where
$w=2,\gamma=\frac{\pi}{2},\alpha=\frac{\pi}{3},D=0.5$. }
	\label{LIN0}
\end{figure}

In this section, we investigate the behavior of the non-local information $\mathcal{I}_{non}$ that coded on all the possible partition,$\rho_{AB}, \rho_{AC}$ and $\rho_{BC}$. There is a possibility of exchanging the non-local information between the three qubits. In Fig.(\ref{LIN0}a), we quantify the amount $\mathcal{I}_{non}$  for the initial state $\rho_{AB}(0)$,  as a function on  the weight parameter $\kappa$ and the angle $\alpha$.  At fixed value of $\alpha=\pi/3$, the amount of the non-local information increases gradually to reach its maximum value, namely $\mathcal{I}_{non}=2$, at $\kappa=1$. Due to the interaction with the control parameter, the density operator which describe the three  qubits is given by $\rho_{ABC}$. Fig.(\ref{LIN0}b) displays the behavior of the amount of the non-local information, $\mathcal{I}_{non}(\rho_{ABC})$, at  $Dz=0.5$ and the dipole strength $\omega=2$. Again as it exhibited from (\ref{LIN0}b), the non-local information of the whole system increases gradually  as $\kappa$ increases to reach its maximum value $\mathcal{I}_{non}(\rho_{ABC})=3$ at $\kappa=1$.  The $DM$  strength will exhibit a change on the behavior of information that could be generated or loses between the three qubits.

\begin{figure}[h!]
	\centering
	\includegraphics[width=0.4\linewidth, height=5cm]{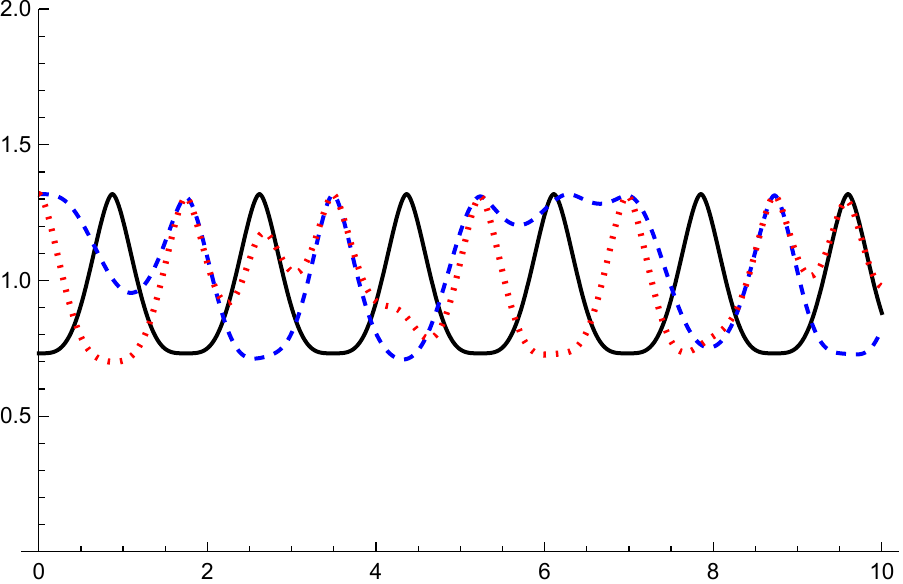}~~\quad
   \put(-70,-10){$t$}
   \put(-50, 130){$(a)$}
\put(-200,50){\rotatebox{90}{\small{$I_{non}(\rho_{ij})$}}}
 \hspace{1.5cm}
\includegraphics[width=0.4\linewidth, height=5cm]{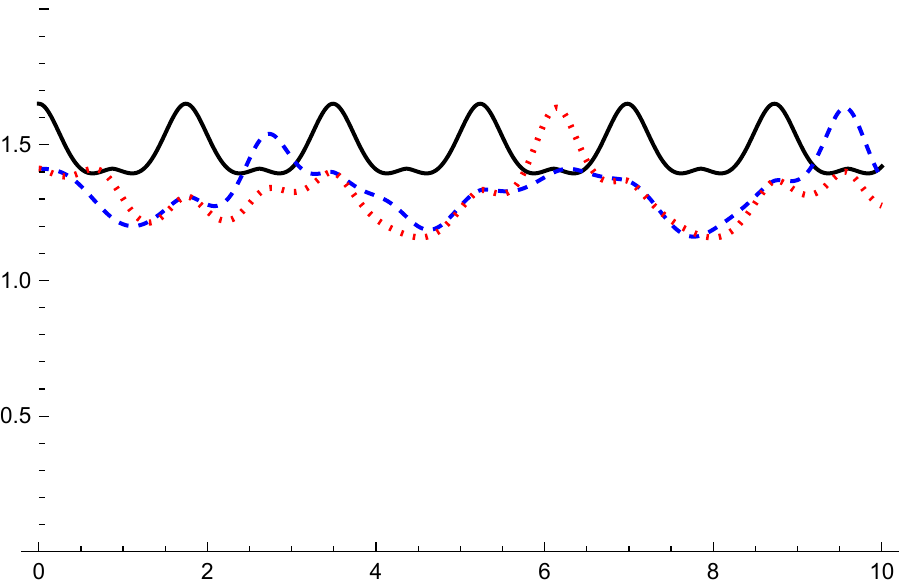}
 \put(-70,-10){$t$}
    \put(-190,50){\rotatebox{90}{\small{$I_{non}(\rho_{ij})$}}}
    \put(-60, 130){$(b)$}
		\caption{The  weight effect of the initial state  on the generated non-local information, where  $D_z=0.9,w=0.5,\gamma=\frac{\pi}{2},\alpha=\frac{\pi}{3}$. The solid(black), dash dot(blue), and the dot (red) curves, represent the non-local information for the states $\rho_{AB}, \rho_{AC}$, and $\rho_{BC}$, respectively where, (a) $\kappa=0.3$, and (b)$\kappa=0.9$.}
	\label{Fig10}
\end{figure}
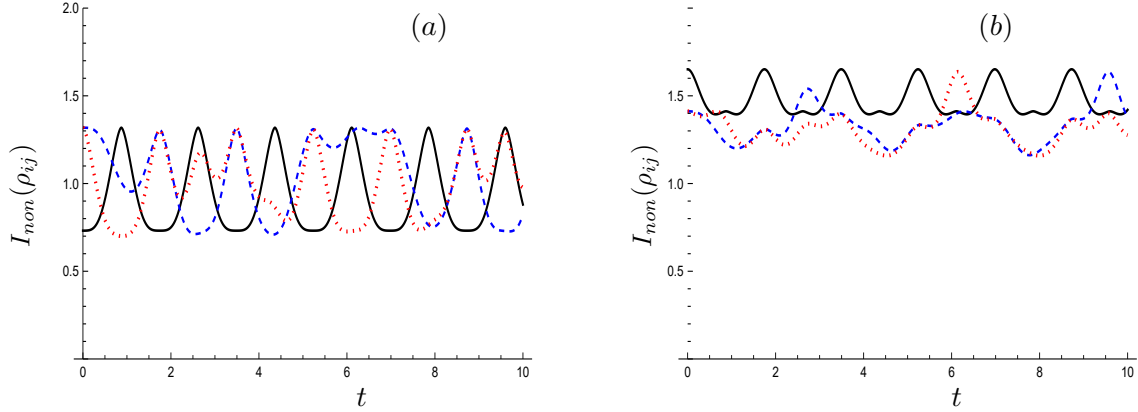

Fig.(\ref{Fig10}) displays the exchange of the non-local information between the three  states, $\rho_{ij}, ij=AB, AC$ and $BC$  at different values of the weight parameter $\kappa$, while the  strength of DM interaction is fixed such that $D_z=0.9$. In Fig.(\ref{Fig10}a), we  investigate the behavior of $\rho_{ij}$ where it is assumed that, the wight parameter $\kappa=0.3$. It is clear that, at $t=0$, the non-local information that is coded in the state $\rho_{AB}$ increases gradually to reach its maximum bounds. However,  as the interaction time $t$ increases the nonlocal information $\mathcal{I}_{AB}$ decreases gradually to its minimum value. This behavior is repeated periodically  at further $t$, where the maximum/minium values of the non-local information are similar during the time's interaction. The similar periodic behavior is predicted for $\mathcal{I}_{non}(\rho_{AC})$  and $\mathcal{I}_{non}(\rho_{AC})$, where as $\mathcal{I}_{AB}$ increases, the non-local information that coded on $\rho_{AC}$ and $\rho_{BC}$ decreases. Whenever, $\mathcal{I}_{AB}$ reaches its maximum values, both $\mathcal{I}_{non}(\rho_{AC})$ and  $\mathcal{I}_{non}(\rho_{BC})$ are minimum. As an important observation, the maximum values of the non-local information  that predicted for $\rho_{AC}$ and $\rho_{BC}$ doesn't  exceed that coded on the initial state $\rho_{AB}$.
The amount of non-local information that is coded on the states $\rho_{AC}$ and $\rho_{BC}$ via direct and indirect interaction by  DM, respectively, oscillates periodically between their  maximum and lower bounds. Their minimum values  don't  exceed the minimum values of $\mathcal{I}_{non}(\rho_{AB})$.
In Fig.(\ref{Fig10}b), we investigate the behavior of the non-local information that coded on the states $\rho_{ij}$ at larger values of the weight parameter, where we set $\kappa=0.9$. A similar behavior is predicted as that shown (\ref{Fig10}a), but the upper and lower bounds of $\mathcal{I}_{non}(\rho_{ij})$ are much better than those shown  at $\kappa=0.3$.

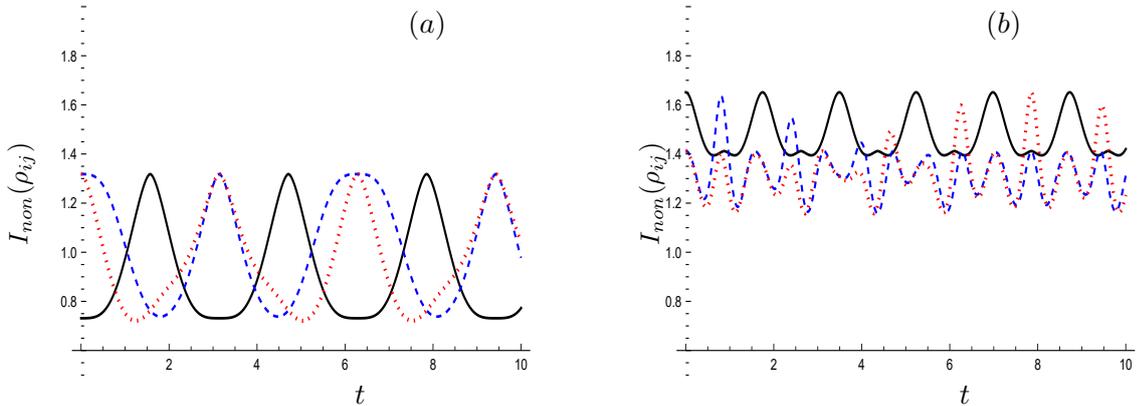
\begin{figure}[h!]
	\centering
\includegraphics[width=0.4\linewidth, height=5cm]{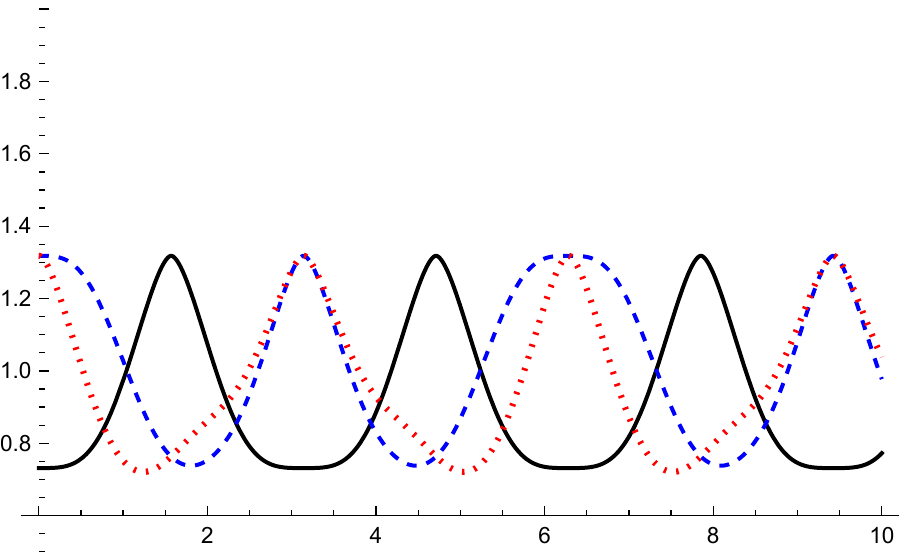}~~\quad
   \put(-70,-10){$t$}
   \put(-50, 130){$(a)$}
\put(-200,50){\rotatebox{90}{\small{$I_{non}(\rho_{ij})$}}}
 \hspace{1.5cm}
\includegraphics[width=0.4\linewidth, height=5cm]{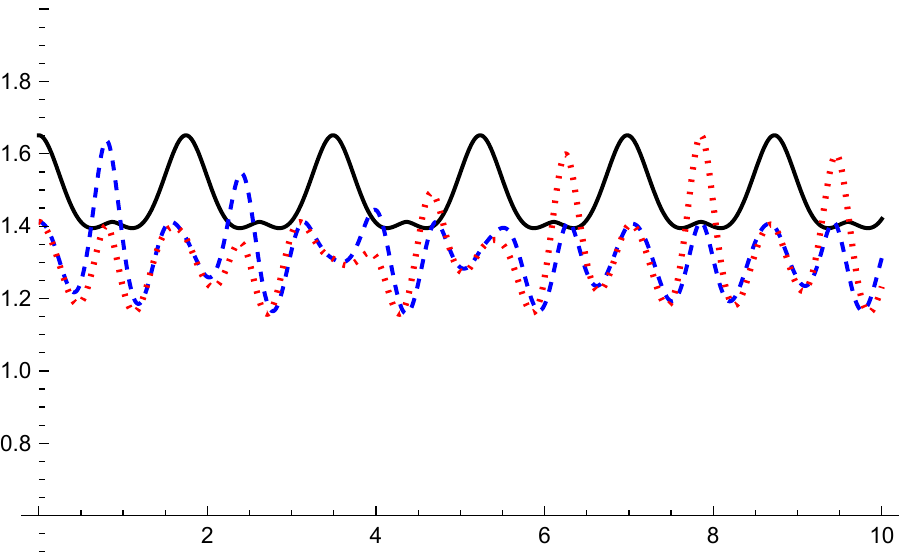}~~\quad
 \put(-70,-10){$t$}
    \put(-190,50){\rotatebox{90}{\small{$I_{non}(\rho_{ij})$}}}
    \put(-60, 130){$(b)$}
		\caption{(a) The same as Fig.(\ref{Fig10}a) but $D_z=0.5$, and (b)is the same as Fig.(\ref{Fig10}b) but with $\omega=2$. }
	\label{Fig11}
\end{figure}

Fig.(\ref{Fig11}a) displays the behavior $\mathcal{I}_{non}(\rho_{ij})$ at smaller values of DM strength, where we set $D_Z=0.5$. It is clear that, the for $\rho_{AC}$ (direct interaction) and $\rho_{BC}$ (indirect interaction)  the non-local information decreases, while the non-local information on the marginal state $\rho_{AB}$ increases. However, the decreasing rate of $\mathcal{I}_{non}$  that depicted for  $\rho_{AC}$ is larger than that displayed in  Fig.(\ref{Fig10}a). Moreover, the crests and the  troughs  of the marginal  states are exchanged between the states $\rho_{AC}$, $\rho_{BC}$ and the initila state $\rho_{AB}$. The effect  of larger values of $\omega$ is shown in Fig.(\ref{Fig11}b),  where we et $\omega=2$.  The behavior is similar to that displayed in Fig.(\ref{Fig10}b). However, the numbers of oscillations are larger, while their amplitudes are smaller, namely the non-local information is better than that shown in Fig.(\ref{Fig10}b).

\section{conclusion}
In this manuscript, a system of two qubits is initially prepared in a partial entangled state governed by $XX$ chain. One of its subsystems interacts locally with a control qubit via Dzyaloshinskii-Moriya (DM). Due to these interactions, there well be entangled states are generated between the three qubits. The possibility of exchanging  the quantum correlations and the non-local information between all  the partitions is discussed. We investigate the effect of the initial state settings and  the  strengths of the interaction on the this process.

It is shown that, at small values of the weight parameter the  ability of DM interaction to generate quantum correlations between the initial two  qubits  is larger than that may be generated by the dipole interaction. However,  large weight parameter is a guarantor for generating long-lived quantum correlation at small strength of DM interaction.  This behavior is changed  if one increases the strength of DM, where the upper bounds are larger while the minimum bounds are smaller than those displayed at small weight parameter. Moreover, the long-lived quantum correlations  are displayed as one increases the strength of dipole interaction.
The numerical computations exhibit that, the quantum correlations  are exchanged  between the initial state the two marginal states, where as quantum correlations decrease on the initial system, it increases on the two marginal systems and vis versa.  The maximum bounds of the quantum correlations that predicted for the marginal systems never exceed that displayed for  the initial system. As soon as quantum correlation of the initial state  vanishes, the marginal states exhibit a maximum quantum correlation.  However, at large  values of interaction strength, one finds that, the quantum correlation of  marginal states have different phase due to the direct/indirect interaction with DM.

The phenomena of the exchanging the non-local information between all the three partitions is examined at different values of the interaction's strength and the weight parameter. Our results display that, the non-local information behaves simultaneously with the quantum correlations, where the large quantum correlations, the large non-local information.  One can maximize the  amount of the non-local information by increasing the strength of DM interaction and decreasing the dipole interaction strength.  However, at large dipole strength, the number of oscillations and their amplitudes increase, and  consequently, the lower bounds of the non-local information decreases.

The generated quantum correlations are quantified by using the concurrence, entanglement of formation and negativity. The behavior of the three quantifiers is discussed, where  the  large amount of quantum correlations are predicted by concurrence and the negativity, while the entanglement of formation displays the smallest values of quantum correlations.


{\it In conclusion}, it is possible to exchange the quantum correlations and the non-local information  between the all  marginal states, which may be generated  via direct or indirect DM interaction. The maximum amount of these correlations and information do not exceed the initial ones. The  interaction parameters are considered as  control parameters for generating long-lived quantum correlations.  We believe that these results are important for generating entangled  quantum network,  since one can generate  long- live entanglement between distant particles as  members of a quantum network and exchanging the non-local information between its members is possible.

‏	

%
%
%
%

\end{document}